\documentclass[aps,pre,twocolumn,groupedaddress,superscriptaddress,showpacs]{revtex4-1}
\usepackage{amsmath}

\usepackage[utf8]{inputenc}
\usepackage[portuguese,english]{babel}

\usepackage{graphicx}
\usepackage{dcolumn}
\usepackage{bm}

\usepackage{cases} 
\usepackage{empheq}

\begin{document}

\title{Optimal number of linkers per monomer in linker-mediated aggregation}

\author{G. C. Antunes} 
\email{antunes@is.mpg.de}
\affiliation{Centro de Física Teórica e Computacional, Universidade de Lisboa, 1749-016 Lisboa, Portugal.}

\author{C. S. Dias}
\email{csdias@fc.ul.pt}
\affiliation{Centro de Física Teórica e Computacional, Universidade de Lisboa, 1749-016 Lisboa, Portugal.}
\affiliation{Departamento de Física, Faculdade de Ciências, Universidade de Lisboa, 1749-016 Lisboa, Portugal.
}%

\author{M. M. Telo da Gama}
\email{mmgama@fc.ul.pt}
\affiliation{Centro de Física Teórica e Computacional, Universidade de Lisboa, 1749-016 Lisboa, Portugal.}
\affiliation{Departamento de Física, Faculdade de Ciências, Universidade de Lisboa, 1749-016 Lisboa, Portugal.
}%

\author{N. A. M. Araújo}
\email{nmaraujo@fc.ul.pt}
\affiliation{Centro de Física Teórica e Computacional, Universidade de Lisboa, 1749-016 Lisboa, Portugal.}
\affiliation{Departamento de Física, Faculdade de Ciências, Universidade de Lisboa, 1749-016 Lisboa, Portugal.
}%

\begin{abstract}
We study the dynamics of diffusion-limited irreversible aggregation of
	monomers, where bonds are mediated by linkers. We combine kinetic Monte
	Carlo simulations of a lattice model with a mean-field theory to study
	the dynamics when the diffusion of aggregates is negligible and only
	monomers diffuse. We find two values of the number of linkers
	per monomer which maximize the size of the largest aggregate. We
	explain the existence of the two maxima based on the distribution of
	linkers per monomer. This observation is well described by a simple
	mean-field model. We also show that a relevant parameter is the ratio
	of the diffusion coefficients of monomers and linkers. In particular,
	when this ratio is close to ten, the two maxima merge at a single maximum.
\end{abstract}

\maketitle

\section{Introduction \label{sec.intro}}
The growth of large structures from the spontaneous aggregation of individual
constituents (monomers) is a subject of interest across fields and disciplines
\cite{Hobbs1974,Zayas2007,Senstad1989,Karanis2002,Do2015,Ruzicka2011,DelasHeras2011,Dobnikar2013,Elliott2003,Geerts2010,
Gnan2012,Nykypanchuk2008,Park2001,Sciortino2004,Singh2015,DelasHeras2010,Zaccarelli2007,Levine2006}.
From the nucleation and growth of crystallites at the nanoscale, flocculation
and self-assembly of colloidal suspensions at the micron scale, to the
formation of social groups at the human scale, there are many examples where
the dynamics of aggregation has been described by simple mechanisms such as
diffusion and reaction/bond formation
\cite{Evans2006,Witten1981,Puljic2005,Araujo2015,Dias2013,Dias2017,Dias2018,Tavares2018,Tavares2010a,Joshi2016, Maye2006}. 

Most of the previous studies considered monomer-monomer bonds that are either
independent or activated by enzymes, where each enzyme may activate more than
one bond~\cite{Margolin1987,Dordick1987,Patil1991,Bisht1997}. Recently, however, the
interest has shifted towards monomer-monomer bonds mediated by a second
species, the
linkers~\cite{Bharti2014,Chen2015,Cyron2013,Lindquist2016,Luo2015,Muller2014,Peng2016}.
Linkers are different from enzymes, for they mediate at most one bond. The idea
is to control aggregation through the properties of the linkers (e.g., their
size, shape, chemistry, and concentration), keeping the properties of
individual monomers intact. But linkers provide many more control parameters to
the dynamics. The dependence on each one of them is still elusive.

The diffusion of aggregates is expected to decrease with the size of the aggregates. Here, to characterize the dependence of the aggregation dynamics on the number of linkers per monomer through their relative coefficients, we considered that the diffusion of aggregates is negligible, when compared to that of monomers. This approach is expected to be meaningful for very diluted regimes as those considered in this work. In fact, this strategy was used before to provide insight into the role of different parameters in the dynamics of aggregation on crystalline substrates \cite{Pai1997,Ehrlich1991} and may be relevant for large molecules and colloidal particles in physiological and porous media \cite{Papadopoulos2000,Li2009a}.
We consider the limit of irreversible
aggregation, not only for theoretical simplicity but also due to its practical
relevance~\cite{Zhang2015,Aran2010}. In many applications, strong irreversible bonds are instrumental to yield resilience to thermal fluctuations and mechanical perturbations~\cite{Yount2005}. For example, our model may be relevant to describe self-assembly through DNA linkers~\cite{Xiong2009}, where the interactions can be very strong (up to 6$k_BT$ per linker~\cite{Biancaniello2005}). Here the length of the linkers may also hinder the diffusion of the clusters due to their geometry~\cite{Kraft2013}.

Numerical simulations reveal two
optimal numbers of linkers per monomer for which the average size of the
aggregates is maximized. We propose a mechanism responsible for this effect
and show that the results may be described by a mean-field calculation.
We also discuss how the existence and the values of the maxima depend on the
diffusion coefficients of the two species.

We introduce the model and simulation details in Sec.~\ref{sec.model}, present
the main results in Sec.~\ref{sec.results}, and draw some conclusions in
Sec.~\ref{sec.conclusion}.

\section{Model \label{sec.model}}
\begin{figure}
\includegraphics[width=\columnwidth]{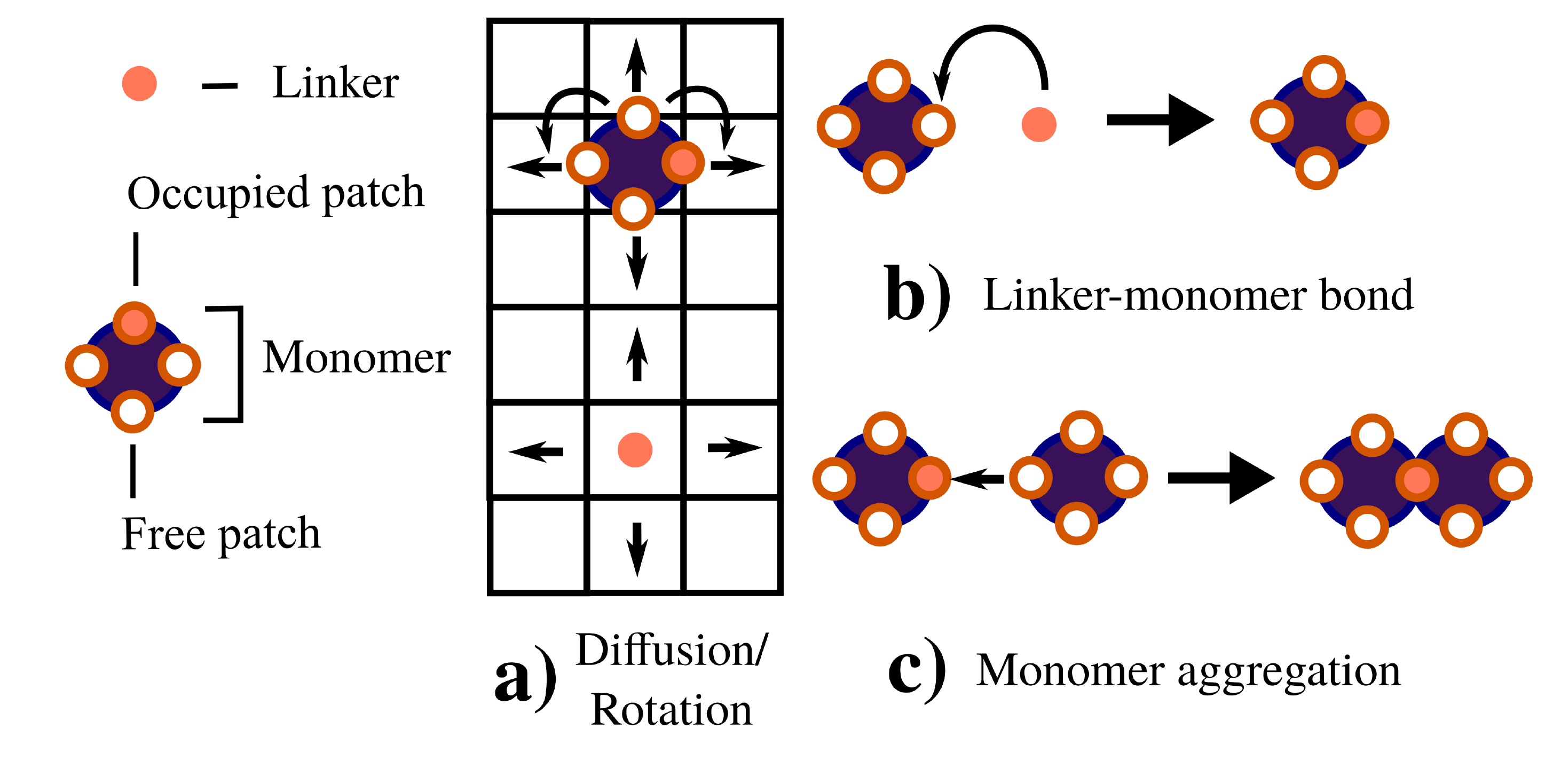}
\caption{Schematic representation of the main events: a) Translational
	(linkers and monomers) and rotational (monomers only) diffusion; b)
	formation of bonds between monomers and linkers; c) formation of
	linker-mediated bonds between monomers.~\label{fig.illustration}}
\end{figure}

Let us consider a cubic lattice with a diluted mixture of two species:
monomers and linkers. Monomers occupy one lattice site and move by a sequence of
random hops between nearest neighboring sites. The hopping rate is equal to
the translational diffusion coefficient $D_\mathrm{m}$. A lattice site cannot be
occupied by more than one monomer. The initial concentration of monomers $n_\mathrm{m}$
is defined as the fraction of lattice sites occupied by monomers at $t=0$.
Monomers have a number $f$ of patches on their surface with $f$ and the
symmetry of their spatial distribution given by the topology of the lattice.
For example, on the cubic lattice considered here, $f=6$ and the patches are
oriented along the three directions of the nearest neighbors. A patch is occupied, when a bond to a linker is established and free, otherwise. Each monomer also rotates $\pm\pi/2$ at a rate equal to the rotational diffusion coefficient $D_\mathrm{r}$ (see Fig.~\ref{fig.illustration}a). A lattice site is rescaled to the monomer's diameter and sets the length scale. Then using the Stokes-Einstein-Debye relation, we consider, for simplicity, $D_\mathrm{r}\approx D_\mathrm{m}$.

The linkers also occupy one lattice site and hop with a diffusion coefficient
$D_\mathrm{l}$. A lattice site can be occupied by at most one linker. The concentration
of linkers $n_\mathrm{l}$ is defined analogously to $n_\mathrm{m}$. Linkers form bonds only to
patches, turning a free patch into an occupied patch (see
Fig.~\ref{fig.illustration}b). By forming at most two bonds, linkers mediate
bonds between monomers (see Fig.~\ref{fig.illustration}c). We define an
aggregate as a set of monomers connected by linker-mediated bonds.  In this
work, we focus on the limit where the diffusion of monomers is much faster than
that of aggregates, such that the latter may be neglected. We also consider
that all bonds are irreversible, i.e., there is no bond breaking in the
timescale of interest.

We consider that the formation of a bond is much faster than all other processes and therefore instantaneous within the relevant time scale. To follow the kinetics, we performed kinetic Monte Carlo simulations of the three relevant processes: translational and rotational diffusion of monomers and translational diffusion of linkers (see Fig.~\ref{fig.illustration}). At each iteration, the process is drawn at random from the list of possible processes, with the probability $n_i W_i/R$ of each process $i$, where $n_i$ is the number of particles that can undergo process $i$, $W_i$ is the rate of the process $i$, and R is the sum over all $n_i W_i$ (normalization). The time is incremented by 1/R.

\section{Results \label{sec.results}}
The model described above has four relevant parameters: the concentrations of
monomers and linkers ($n_\mathrm{m}$ and $n_\mathrm{l}$, respectively) and their diffusion
coefficients ($D_\mathrm{m}$ and $D_\mathrm{l}$, respectively). For simplicity, time is rescaled such that the diffusion coefficient of monomers is unity, reducing the latter two parameters to a single adimensional one, the ratio of diffusion coefficients, defined as,
\begin{equation}
	\Delta = \frac{D_\mathrm{m}}{D_\mathrm{l}},
\end{equation}
where, for instance, when $\Delta=10^{-3}$, the linkers undergo 1000 diffusion steps while the monomers undergo a single one.

Below, we investigate first the limit of fast linkers, where
$\Delta\rightarrow0$, and then generalize the study to any value of $\Delta$.

\subsection{Fast linkers ($\Delta\rightarrow0$) \label{sec.results.separate}}
\begin{figure}[t]
\includegraphics[width=\columnwidth]{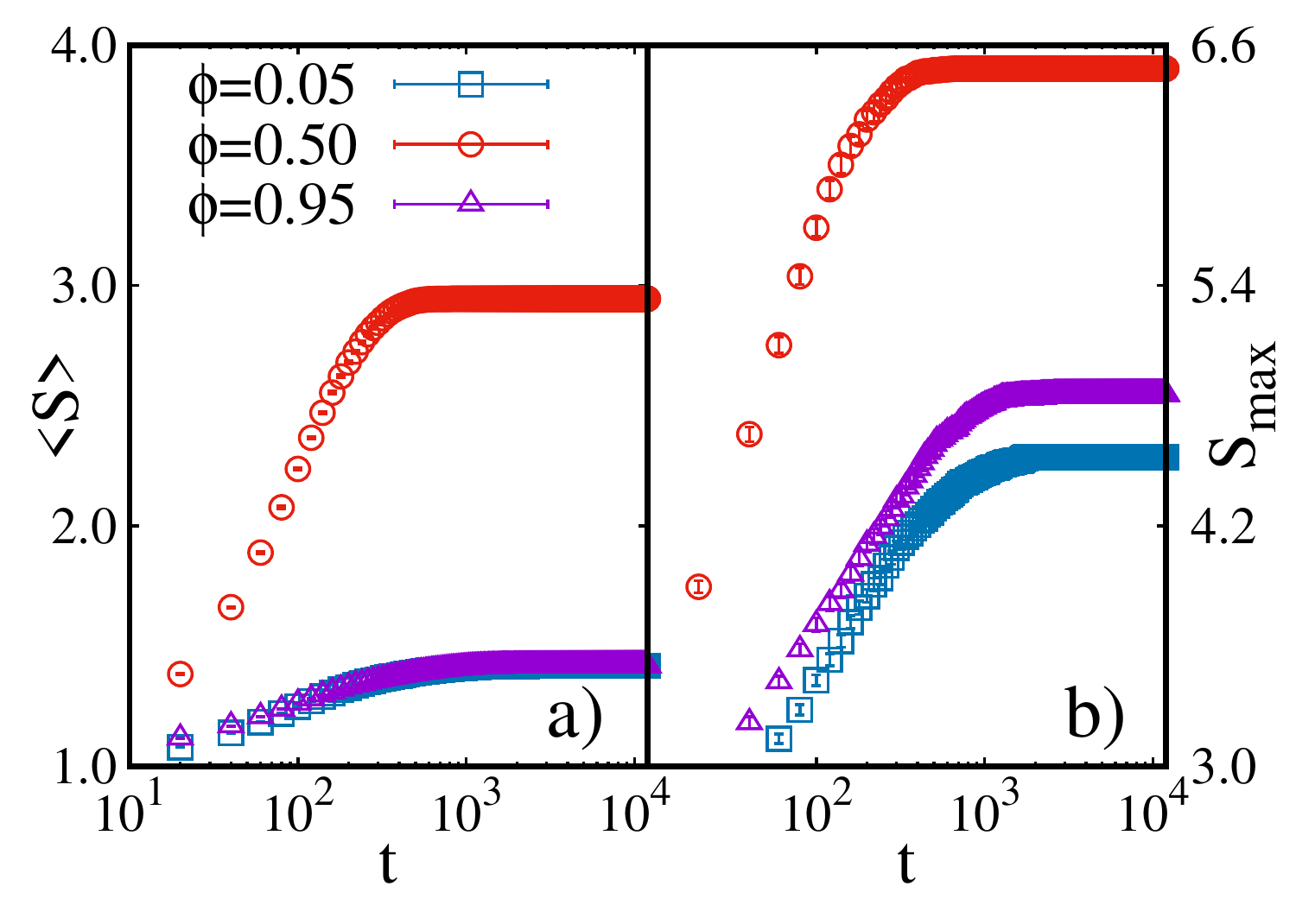}
	\caption{a) Time evolution of the average size of the aggregates
	$\langle S \rangle$, defined in Eq.~\eqref{eq.S_definition}. b) Time
	evolution of the size of the largest aggregate $S_\mathrm{max}$.
	Simulations were performed on a cubic lattice of lateral size
	$L_\mathrm{box}=25$ in units of the lattice constant, for a ratio of monomer
	and linker diffusion coefficients $\Delta=10^{-3}$, and a number of
	linkers per patch of $\phi=\{0.05,0.50,0.95\}$. Results are averages
	over $500$ samples.~\label{fig.timeEvol}}
\end{figure} 

\begin{figure}[t]
	\includegraphics[width=\columnwidth]{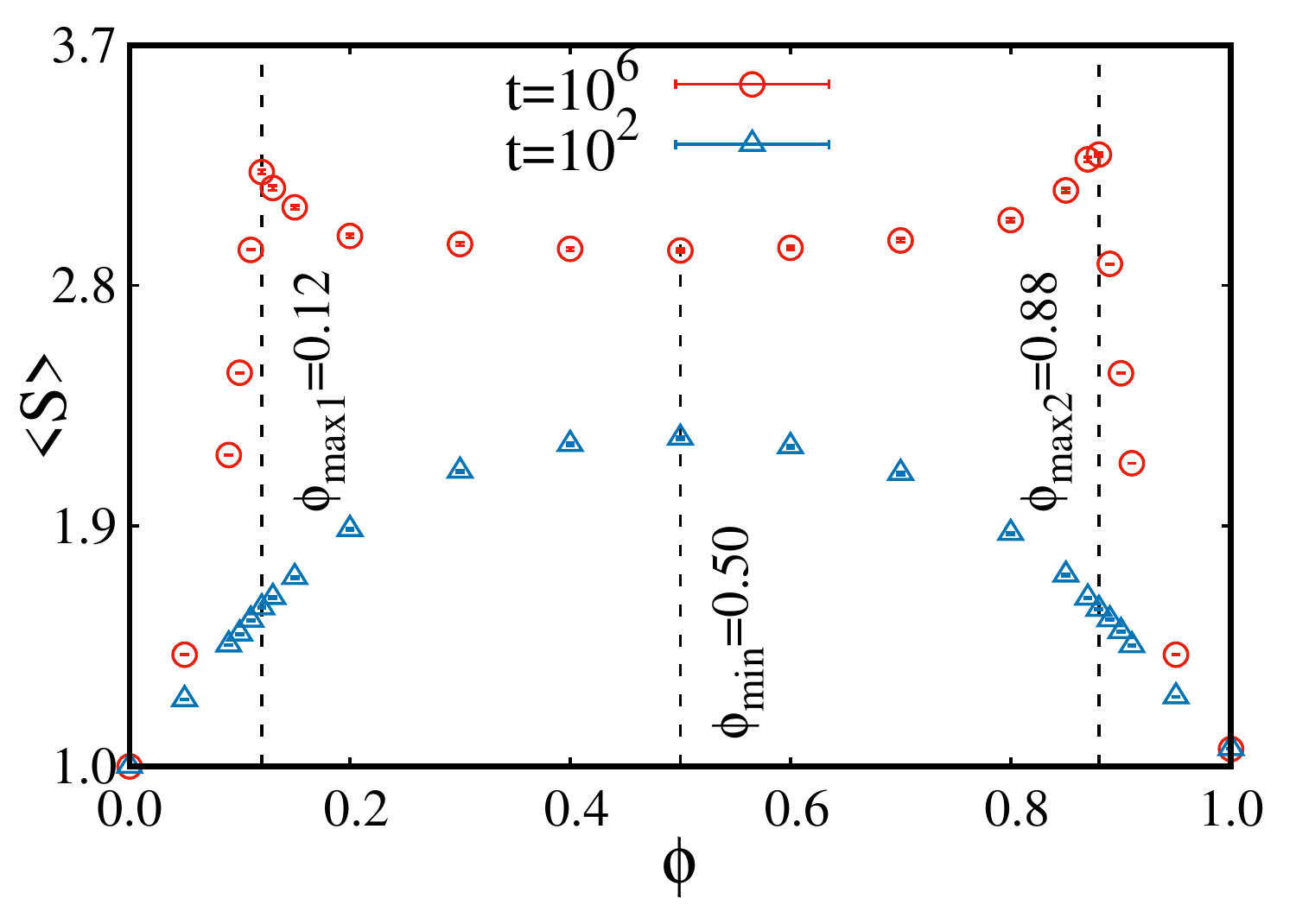}
	\caption{Average size of the aggregates $\langle S \rangle$ as a
	function of $\phi$, at two values of the time: $10^2$ and $10^6$, in units
	of $a^2/D_\mathrm{m}$, where $a$ is the lattice constant and $D_\mathrm{m}$ the diffusion
	coefficient of monomers.  The dashed (vertical) lines mark the two
	maxima and one minimum, observed at $\phi=\{0.12, 0.50, 0.88\}$.
	Simulations were performed on a cubic lattice of lateral size
	$L_\mathrm{box}=25$ in units of the lattice constant, for a ratio of monomer
	and linker diffusion coefficients of $\Delta=10^{-3}$, and a
	concentration of monomers of $n_\mathrm{m}=0.01$. Results are averages over
	$500$ samples.~\label{fig.bimodal}}
\end{figure}

Let us first consider the limit where linkers are much faster than monomers,
i.e.,  $\Delta\rightarrow0$. We define the size of an aggregate as the number
of monomers in it. In Fig.~\ref{fig.timeEvol} is the time evolution of the size
of the largest aggregate $S_\mathrm{max}$ and the average size of the
aggregates $\langle S \rangle$, obtained numerically for $\Delta=10^{-3}$.
$\langle S\rangle$ is defined as
\begin{equation}\label{eq.S_definition}
	\langle S \rangle = \frac{N}{N_\mathrm{agg}+N_\mathrm{mon}},
\end{equation}
where $N$ is the total number of monomers, $N_\mathrm{agg}$ is the number of
aggregates, and $N_\mathrm{mon}$ is the number of free monomers, i.e., monomers
without linker-mediated bonds. Note that for $\langle S\rangle$ we considered
also the monomers.  Both parameters increase monotonically in time and saturate
asymptotically, for the following reason. As monomers and linkers find each
other through diffusion, free patches become occupied. When an occupied patch
finds a free one, they form a linker-mediated bond between the corresponding
monomers. For irreversible aggregation, the number of free monomers
$N_\mathrm{mon}(t)$ is a decreasing function of time. Both the average size of the
aggregates $\langle S \rangle$ and the size of the largest aggregate
$S_\mathrm{max}$ increase in time and they saturate when all possible bonds are
formed.  The asymptotic values of $S_\mathrm{max}$ and $\langle S\rangle$
depend strongly on the number of linkers per patch $\phi$,
\begin{equation}
	\phi = \frac{n_\mathrm{l}}{f n_\mathrm{m}} \ \ ,
\end{equation}
as shown in Fig.~\ref{fig.bimodal}. While for $\phi=0.05$ all linkers form
bonds to patches in the asymptotic limit, as there are more patches than
linkers, for $\phi=0.95$ the aggregation ceases long before all linkers form a
bond. Note that, for $\phi>0.5$, there are more linkers than needed for each
patch to form a bond.

Figure~\ref{fig.bimodal} shows the dependence of $\langle S \rangle$ 
on $\phi$ for two values of the time. At the earliest time (triangles), $\langle
S\rangle$ is maximal for $\phi=0.5$, as one would expect, for it corresponds to
an equal amount of free and occupied patches and therefore a maximal number of
possible bonds. However, at later times (circles), one finds two maxima
instead (for $\phi_\mathrm{max1}=0.12$ and $\phi_\mathrm{max2}=0.88$) and $\phi=0.5$ is in
fact a local minimum.  We performed simulations for different box sizes, namely,
$L_\mathrm{box} \in \{16,25,32,64\}$, and observed no significant finite-size
effects.

\begin{figure}[t]
	\includegraphics[width=\columnwidth]{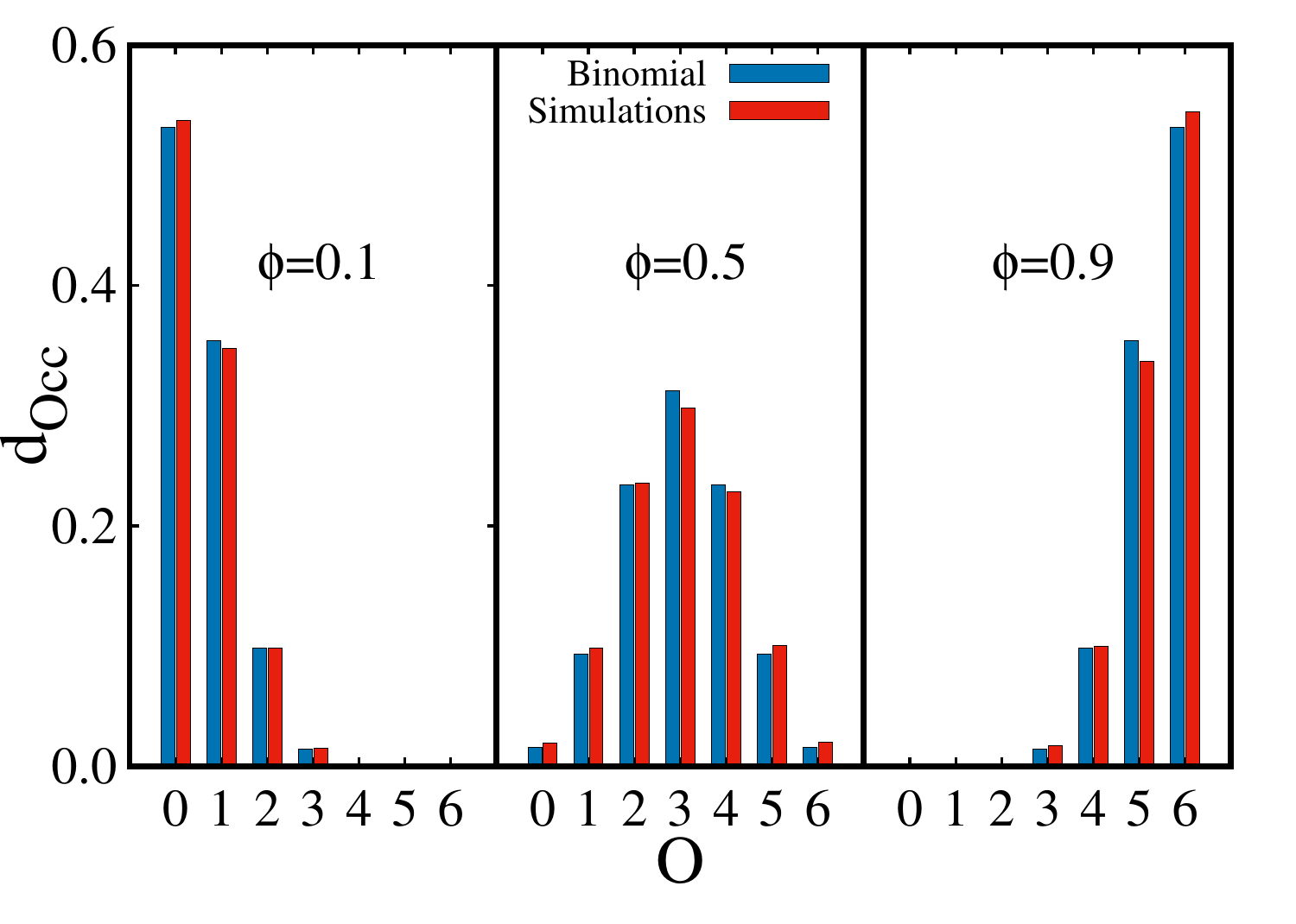}
	\caption{Probability distribution of the number number of occupied
	patches per monomer obtained from numerical simulations (red) and from
	the binomial distribution (blue), given by Eq.~\eqref{eq.binomial}.
	The numerical simulations were performed on a cubic lattice of lateral
	size $L_\mathrm{box}=25$, in units of the lattice constant, suppressing the
	diffusion of monomers ($\Delta=0$) for a concentration of monomers of
	$n_\mathrm{m}=0.01$, as in Fig.~\ref{fig.bimodal}. Results are averages over
	$500$ samples.~\label{fig.distOcc}}
\end{figure}
To understand the dependence on $\phi$, let us focus on the limit where linkers
are infinitely faster than monomers $\Delta=0$, i.e., when all possible
linker-monomer bonds are formed before the diffusion of monomers starts to play
a role. Thus, the relevant parameter for monomer-monomer aggregation is the
probability $d_\mathrm{Occ}(O)$ that a monomer has $O$ occupied patches, after all
possible linker-monomer bonds are formed. If we neglect spatial correlations,
this probability is given by the binomial distribution,
\begin{equation}\label{eq.binomial}
	d_{Occ}(O)=\binom{f}{O}\phi^O (1-\phi)^{f-O} \ \ ,
\end{equation}
where $\phi$ is the probability that a given patch is occupied. 
Figure~\ref{fig.distOcc} depicts the probability distributions $d_\mathrm{Occ}(O)$ obtained
from Eq.~ \eqref{eq.binomial} and from numerical simulations at the same value
of the concentration of monomers considered in Fig.~\ref{fig.bimodal} but with
$\Delta=0$, showing that for this concentration, spatial correlations are
practically negligible. 
\begin{figure}
	\includegraphics[width=\columnwidth]{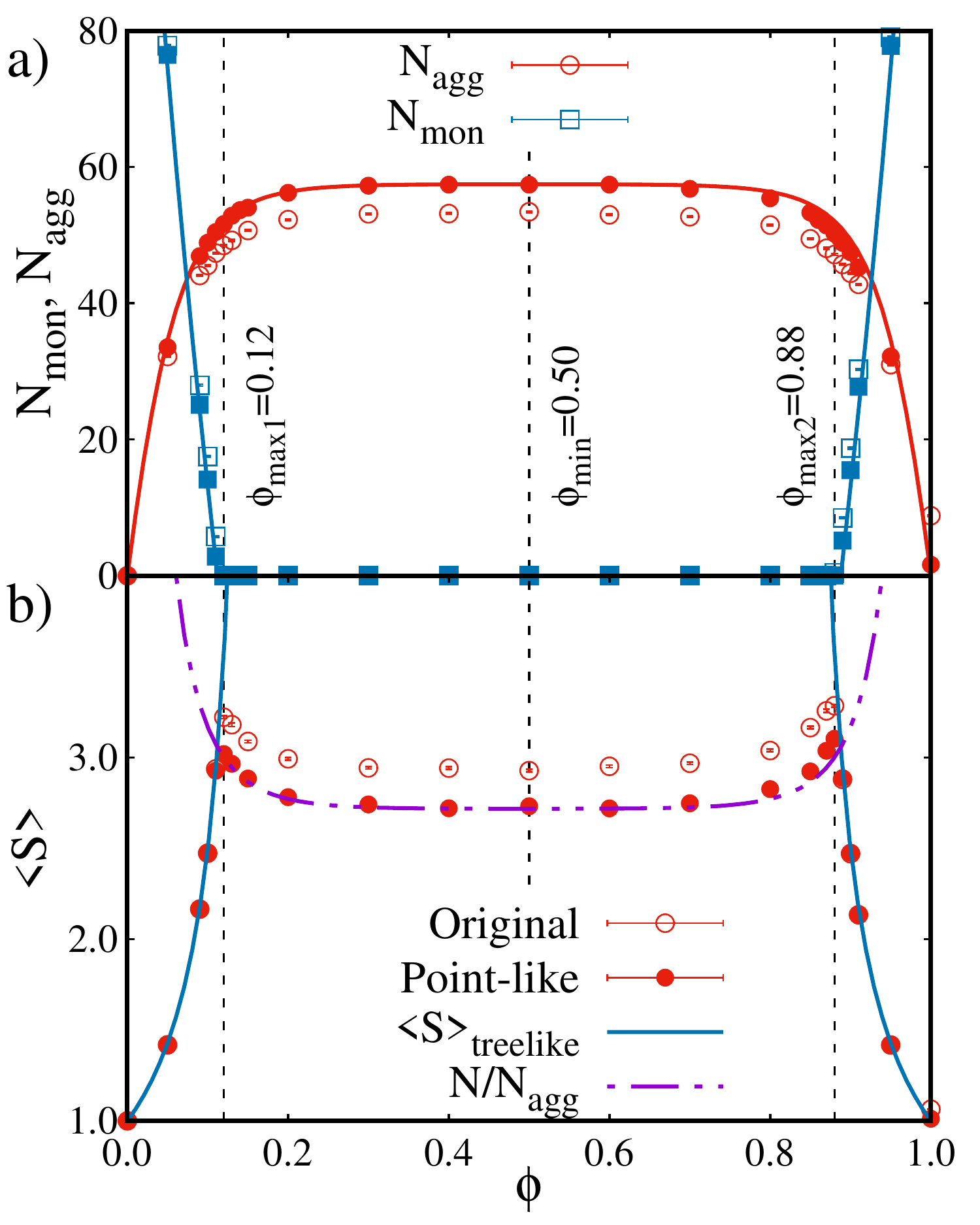}
	\caption{a) Number of aggregates $N_\mathrm{agg}$
	and number of free monomers $N_\mathrm{mon}$ in the asymptotic limit.  
	Lines are obtained from
	Eqs.~\eqref{eq.S_definition},~\eqref{eq.aggDyn0},~\eqref{eq.treeLike1},
	and~\eqref{eq.treeLike2}.
	b) Average size of the aggregates $\langle S \rangle$ as a
	function of $\phi$ in the asymptotic limit. The full line is obtained
	from Eqs.~\eqref{eq.treeLike1} and~\eqref{eq.treeLike2}, and the dashed
	line is obtained from the asymptotic solution of
	Eq.~\eqref{eq.aggDyn0}. In both a) and b), open circles are obtained
	from simulating the original model and filled circles are obtained from
	simulating the point-like model. The dashed (vertical) lines mark the
	local maxima and minimum corresponding to $\phi=\{0.12, 0.50, 0.88\}$.
	Simulations were performed on a cubic lattice of lateral size
	$L_\mathrm{box}=25$, in units of the lattice constant, for a ratio of monomer and
	linker diffusion coefficients $\Delta=10^{-3}$ and a monomer
	concentration $n_\mathrm{m}=0.01$. Results are averages over $500$
	samples.~\label{fig.NMon}}
\end{figure}

Figure~\ref{fig.NMon}(a) shows the dependency on $\phi$ of the asymptotic
number of free monomers $N_\mathrm{mon}$ and number of aggregates $N_\mathrm{agg}$. Three
regimes are clearly seen with boundaries that coincide with the maxima in the
asymptotic $\langle S \rangle$. In the regime of low $\phi$ ($\phi \in
[0,0.12]$), most monomers only have free patches and so the number of bonds
equals that of linkers. Since aggregates are immobile, every new
linker-mediated bond involves one monomer (at least). Thus, the monotonic
decrease of $N_\mathrm{mon}$ is well described by a linear dependence on $\phi$.
Accordingly, $N_\mathrm{agg}$ increases with $\phi$ but, since a free monomer with
occupied patches can either form a bond with another free monomer or with a
larger aggregate, $N_\mathrm{agg}$ is sub-linear in $\phi$. By symmetry, the same is
observed in the regime of large values of $\phi$ ($\phi \in [0.88,1]$), 
where $N_\mathrm{mon}$ increases
and $N_\mathrm{agg}$ decreases with $\phi$, as the number of monomers with all
patches free is $(1-\phi)^f$.

For the intermediate regime, $\phi\in[0.12,0.88]$, all monomers form at least
one bond, i.e., $N_\mathrm{mon}=0$. Thus, the average size of aggregates $\langle
S\rangle$ depends only on the number of aggregates, i.e., $\langle
S\rangle=N/N_\mathrm{agg}$. Since it requires a pair of an occupied and a free patch
to form a bond, a monomer with both free and occupied patches may form bonds
with any other monomer. Instead, monomers with only occupied (or only free)
patches cannot form bonds among each other. Note that, in
Fig.~\ref{fig.NMon}(a), $N_\mathrm{agg}$ is maximized for $\phi=0.5$, which
corresponds to the maximum number of monomers with both occupied and free
patches, as given by Eq.~\eqref{eq.binomial}. Since for $\phi=0.5$, more than
$95\%$ of all monomers can form bonds with each other, there is a tendency to
form dimers. For $\phi=0.12$, almost $50\%$ of the monomers have only free
patches. Since those cannot form bonds among each other, they diffuse for a
long time (on average) until they find a monomer (or an aggregate) with one or
more occupied patches to form a bond with, what favors the growth of larger
aggregates. The positions of the maxima ($\phi=\{0.12,0.88\}$) correspond to
the limiting values for which $N_\mathrm{mon}=0$.

Let us now develop a mean-field approach for the dynamics of the monomers, where we neglect spatial correlations and assume that the diffusion of linkers is infinitely faster than that of monomers. From the distribution of patches occupied by linkers $d_\mathrm{Occ}$, we classify monomers as: full (fully occupied or completely free of linkers) and partial (partially occupied by linkers). We define $N_\mathrm{full}(t)$ as the number of full monomers, and $N_\mathrm{partial}(t)$ as the number of partially occupied monomers. In the mean-field limit, the time evolution of these quantities is given by, 

\begin{subequations}
	\begin{align}
	\label{eq.aggDyn0}
	\dot{N}_\mathrm{agg} &=   q_\mathrm{0} \ N_\mathrm{partial} N_\mathrm{full} + \frac{q_\mathrm{0}}{2} \
		N_\mathrm{partial}^2,\\
	\label{eq.aggDyn1}
	\dot{N}_\mathrm{partial} &= - q_\mathrm{0} \ N_\mathrm{partial} N_\mathrm{full} - q_\mathrm{0} \ N_\mathrm{partial}^2 - q_\mathrm{1} \
		N_\mathrm{partial} N_\mathrm{agg},\\
	\label{eq.aggDyn2}
	\dot{N}_\mathrm{full} &= - q_\mathrm{0} \ N_\mathrm{partial} N_\mathrm{full} - q_\mathrm{1} \ N_\mathrm{full} N_\mathrm{agg},
	\end{align}
\end{subequations}
with initial conditions,
\begin{subequations}
	\begin{align}
	N_\mathrm{agg}(0)  &= 0, \label{eq.aggDynStart0}\\
	N_\mathrm{partial}(0) &= N \sum\limits_{n=1}^{f-1} d_\mathrm{Occ}(n)
		\label{eq.aggDynStart1} ,\\
	N_\mathrm{full}(0) &= N [ \ d_\mathrm{Occ}(0) + d_\mathrm{Occ}(f) \label{eq.aggDynStart2} \
		].
	\end{align}
\end{subequations}
The first term in Eqs.~\eqref{eq.aggDyn0}--\eqref{eq.aggDyn2} is related to
the aggregation between monomers with both occupied and free patches and
monomers with only free or occupied patches. The second term in
Eqs.~\eqref{eq.aggDyn0} and~\eqref{eq.aggDyn1} is related to the aggregation
between monomers with both occupied and free patches. The third term in
Eq.~\eqref{eq.aggDyn1} and the second term in Eq.~\eqref{eq.aggDyn2} are
related to the formation of bonds between monomers and aggregates. 

We consider constant kernels $q_\mathrm{0}$ and $q_\mathrm{1}$. Free monomers form
linker-mediated bonds with each other at a rate $q_\mathrm{0}$ and with aggregates at
rate $q_\mathrm{1}$. Since the latter are considered immobile, then
$q_\mathrm{1}=q_\mathrm{0}/2$~\cite{Chandrasekhar1943}. The absolute value of $q_\mathrm{0}$ sets the
timescale. Since we are only interested in the asymptotic behavior, without
loss of generality, we consider $q_\mathrm{0}=1$.  We neglected also the aggregation
between monomers with only occupied patches and those with only free
patches, for when the number of monomers of one type is significant the number
of the other is negligible.  In Fig.~\ref{fig.NMon}, the lines for $N_\mathrm{agg}$
and $\langle S\rangle$ are obtained numerically from
Eqs.~\eqref{eq.S_definition}~and~\eqref{eq.aggDyn0}, where for
$\phi\in[0.12,0.88]$, we set $N_\mathrm{mon}=0$. The results obtained are
in qualitative agreement with the results from the numerical simulations (open
circles).

To write down Eqs.~\eqref{eq.aggDyn1}~and~\eqref{eq.aggDyn2}, we assumed that
all monomers form bonds. However, this is not the case for the first and the
third regimes. To obtain $\langle S \rangle$ in those cases, we now consider
that aggregates are treelike, such that $N_\mathrm{agg}+N_\mathrm{mon}$ is always reduced
by one when a new bond is formed, i.e., either a dimer is formed or a monomer
is added to a pre-existing aggregate. For $\phi <0.12$ (first regime), all $L$
linkers are mediating a bond. Thus, $N_\mathrm{agg}+N_\mathrm{mon}=N-L$ and
\begin{equation}~\label{eq.treeLike1}
	\langle S \rangle_\mathrm{treelike}=\frac{N}{N-L}=\frac{1}{1-f\phi} \ ,\
	\phi \leq \frac{1}{f},
\end{equation}
which diverges at $\phi=1/f\approx0.17$.  Note that the first regime
stops before that. For $\phi > 0.88$, all patches that are initially free
will participate in a bond. In this case, $N_\mathrm{agg}+N_\mathrm{mon}= N - (Nf - L)$.
In the same way, we obtain,
\begin{eqnarray}
	\langle S \rangle_\mathrm{treelike}=\frac{N}{N-(N f - L)}= \nonumber \\
	\ \ \ \ \ \ \ \ =\frac{1}	{1-f(1-\phi)}  , \frac{f-1}{f}<\phi.
	\label{eq.treeLike2} 
\end{eqnarray}
Figure~\ref{fig.NMon} shows that
Eqs.~\eqref{eq.treeLike1}~and~\eqref{eq.treeLike2} (lines) reproduce
quantitatively the dependence of the average size of the aggregates on $\phi$ for
the first and third regimes. The intersections of $\langle S
\rangle_\mathrm{treelike}$ with the prediction for the second regime, in
Eq.~\eqref{eq.aggDyn0}, coincides with the limits of the different regimes
obtained numerically. Thus, in spite of its simplicity, our mean-field
approach describes qualitatively the average size of the aggregates for the three
different regimes and quantitatively for the first and the third regimes. For
the second, the mean-field approach predicts a larger number of aggregates and
(consequently) smaller aggregates than observed numerically (open circles in
Fig.~\ref{fig.NMon}).

To understand the discrepancies observed in the second regime, we discuss now the role of
the shape of the aggregates. Studies in irreversible aggregation often report
non-compact structures~\cite{Witten1981,Meakin1983}. In general, the shape of
the aggregates should also depend on $\phi$. We now consider a point-like
lattice model, where we neglect the shape of the aggregates, i.e., each
aggregate occupies a single lattice site. Accordingly, every patch in an
aggregate (either free or occupied) is equally likely to form a bond with a
monomer/linker. Thus, when a linker and a monomer or an aggregate try to occupy the
same lattice site while diffusing, they form a bond with a
probability given by,
\begin{equation}
	p_\mathrm{LM}=\frac{F}{F+O},
\end{equation}
where $F$ and $O$ are the number of free and occupied patches of the aggregate,
respectively. Also, when a monomer $i$ hops to a lattice site that is occupied
by an aggregate $j$, they form a bond with a probability,
\begin{equation}
p_\mathrm{MM}=\frac{F_\mathrm{i}O_\mathrm{j} + F_\mathrm{j}O_\mathrm{i}}{(F_\mathrm{i}+O_\mathrm{i})(F_\mathrm{j}+O_\mathrm{j})},
\end{equation}
where $\{F_\mathrm{i},O_\mathrm{i}\}$ and $\{F_\mathrm{j},O_\mathrm{j}\}$ are the number of free and occupied
patches of $i$ and $j$, respectively.  The numbers of free and occupied patches
of the resulting aggregate ($F_\mathrm{final}$ and $O_\mathrm{final}$, respectively) are then
\begin{eqnarray}
	F_\mathrm{final}=F_\mathrm{1}+F_\mathrm{2}-1,\\
	O_\mathrm{final}=O_\mathrm{1}+O_\mathrm{2}-1.
\end{eqnarray}
Note that we consider that all aggregates are treelike (no loops). 

The numerical results for the point-like model are shown also in
Fig.~\ref{fig.NMon} (filled circles). There is a remarkable quantitative
agreement between the point-like and the mean-field calculation. This suggests
that the deviations from the mean-field behavior for the second regime are due
to the size of the aggregates.

\subsection{Dependence on the linker diffusivity \label{sec.results.mixing}}

\begin{figure}[t]
	\includegraphics[width=\columnwidth]{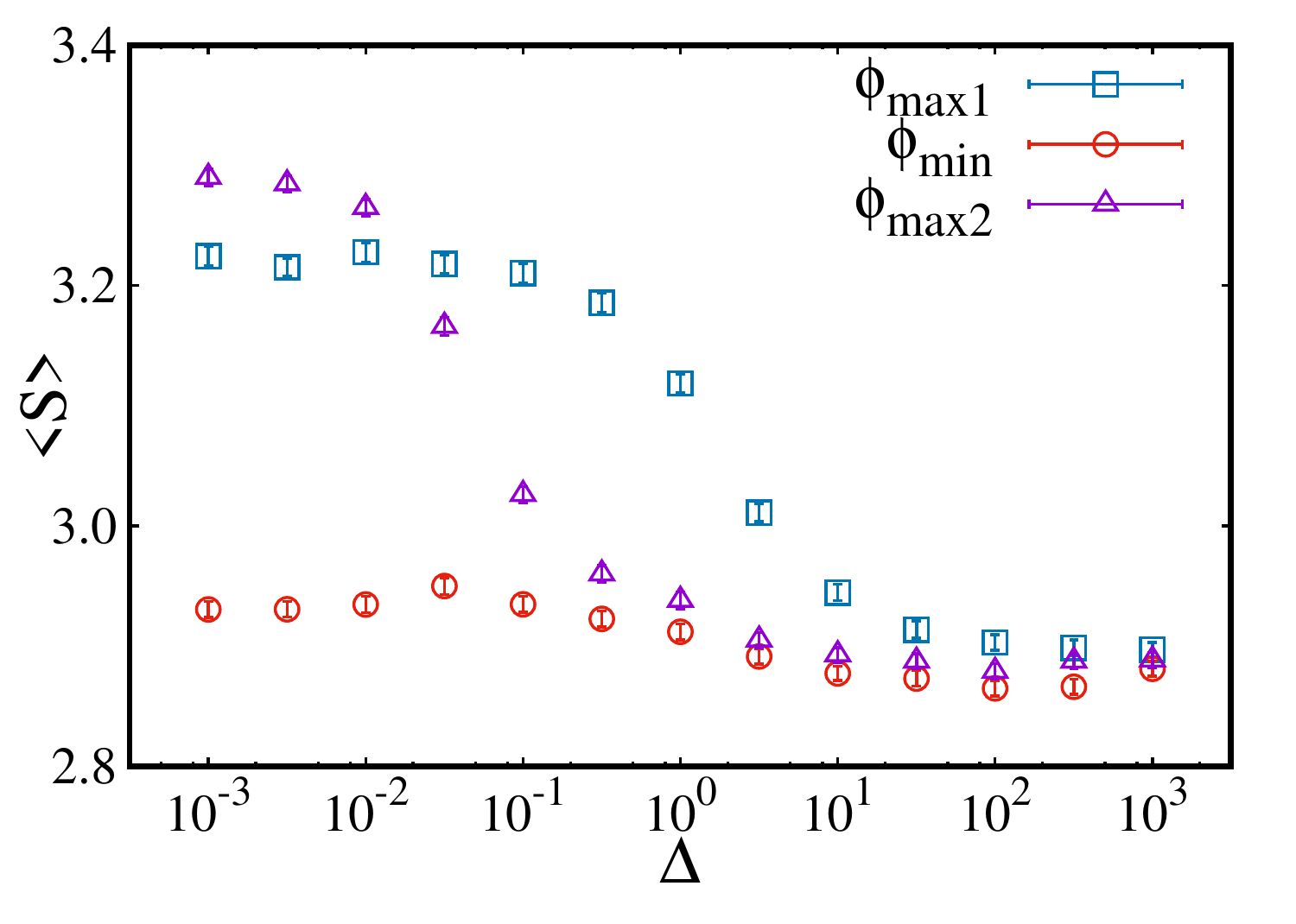}
	\caption{Dependence on the ratio of monomer and linker diffusion
	coefficients $\Delta$ of the asymptotic maxima and minimum of the average size of
	the aggregates $\langle S \rangle$.  Squares correspond to
	$\phi_\mathrm{max1}=0.12$ and triangles to $\phi_\mathrm{max2}=0.88$.  Simulations
	were performed on a cubic lattice of lateral size $L_\mathrm{box}=25$ in
	units of the lattice constant, and a monomer concentration of $n_\mathrm{m}=0.01$.
	Results are averages over $500$ samples.~\label{fig.maxima_delta}}
\end{figure}

So far, we have considered $\Delta\rightarrow0$ and assumed a separation of
timescales between the diffusion of linkers and the one of monomers. However,
above a certain value of $\Delta$, these two mechanisms should compete and
therefore the assumption is no longer reasonable. We now discuss the behavior
of the average size of the aggregates $\langle S\rangle$ for different values
of $\Delta$ in a range covering six orders of magnitude ($10^{-3},10^3$). The
first remarkable observation is that the positions of the two maxima
($\phi_\mathrm{max1}$ and $\phi_\mathrm{max2}$) and that of the minimum ($\phi_\mathrm{min}$) do not vary
significantly with $\Delta$ (not shown). Also, for low values of $\Delta$
($<10^{-2}$), $\langle S\rangle$ is constant and consistent with the mean-field
value (see Fig.~\ref{fig.maxima_delta}). But, as $\Delta$ increases, the
differences between the optimal and the (local) minimum value of $\langle S\rangle$ decrease
and, for $\Delta>10$, they are indistinguishable. 

A monomer-monomer bond is always preceded by the formation of a linker-monomer
bond. As discussed in the previous section, for low values of $\Delta$, the
diffusion of linkers is much faster than the one of monomers, thus all
possible linker-monomer bonds are formed in a timescale that is much faster
than that of the diffusion of monomers. For large values of $\Delta$, since
aggregates are immobile, and monomers are much faster than linkers, it is also
more likely that a free linker forms a bond to a free monomer rather than to
an immobile aggregate. In the same way, it is more likely that the next
monomer-monomer bond leads to the formation of a new immobile aggregate than
to the growth of a pre-existing one. This competition promotes the growth of
the number of aggregates rather than their size. That is why, for all values
of $\phi$ considered in Fig.~\ref{fig.maxima_delta}, we observe a monotonic
decrease of $\langle S\rangle$ with $\Delta$.

In Fig.~\ref{fig.maxima_delta} it is clear that, for $\phi_\mathrm{max1}$, the regime
of fast linkers ($\Delta\rightarrow0$) is valid for a larger range of values of
$\Delta$ than for $\phi_\mathrm{max2}$. In the case of $\phi_\mathrm{max1}$, every linker
mediates a monomer-monomer bond (asymptotically). Thus, it is only when the
diffusion of monomers is comparable to the diffusion of linkers
($\Delta\approx1$) that the formation of linker-monomer bonds competes with the
formation of the next monomer-monomer bond. For $\phi_\mathrm{max2}$, this competition
becomes relevant at lower values of $\Delta$ for the following reason. When
$\Delta\ll1$ (see previous section), all linkers form linker-monomer bonds
promptly and the dynamics of monomer-monomer bonds is controlled by the few
monomers with free patches. Thus, the number of pairs of monomers that can form
a new aggregate is reduced as $\phi$ increases, what promotes the growth of
immobile aggregates. As $\Delta$ increases, the formation of linker-monomer
bonds competes with the one of monomer-monomer bonds. The larger is the value
of $\Delta$, the more likely it is that, a monomer with occupied patches forms
a monomer-monomer bond before all its patches are occupied. This process favors
the formation of new aggregates over the growth of pre-existing ones. 

\section{Conclusion \label{sec.conclusion}}
We studied the dynamics of irreversible aggregation mediated by linkers.  In
the limit of linkers much faster than monomers, we report a bimodal dependence
of the average size of the aggregates on the ratio of the concentrations of
linkers and monomers, with two maxima appearing at non-trivial values. This
behavior appears at long times, being preceded by a transient regime in which
the average size of the aggregates is maximized when the concentration of
linkers is half that of the patches. With a mean-field approach we have shown that
the two maxima appear when the number of free monomers vanishes asymptotically.
This simple approach describes qualitatively and quantitatively the results
obtained numerically. The deviations from the mean-field behavior result from 
effects due to the shape and size of the aggregates that are neglected in the theoretical description.

We investigated also the dependence on the diffusion coefficient of linkers and
monomers. We found that the two maxima disappear when the diffusion of linkers
is comparable to that of monomers (or even lower). This effect results from 
the competition between the formation of linker-monomer and
monomer-monomer bonds.

For simplicity, we considered only irreversible aggregation and immobile
aggregates. However, these two mechanisms are expected to play a role, at
least on much longer timescales. Future studies might consider the effects of both. 
The symmetry of the results is an artefact of the model and in more realistic settings is not expected to occur. 
However, the mechanisms are general and do not rely on a particular symmetry and thus the effects 
reported here are expected to be observed in more realistic, including off-lattice, models and in experiments.  

\section{Acknowledgements}
We acknowledge financial support from the
Portuguese Foundation for Science and Technology (FCT) under Contracts
nos. PTDC/FIS-MAC/28146/2017, LISBOA-01-0145-FEDER-028146, UID/FIS/00618/2019, and SFRH/BPD/114839/2016.

\bibliography{paper}

\end{document}